\documentclass[useAMS,usenatbib,letterpaper]{mn2e}
\usepackage{times}
\usepackage{amssymb}
\usepackage{amsmath}
\usepackage{lscape,graphicx}
\usepackage{float}

\title[High velocity outflows from young star-forming galaxies in the
  UKIDSS Ultra-Deep Survey]{
High velocity outflows from young star-forming galaxies in the
  UKIDSS Ultra-Deep Survey}

\author[E. J. Bradshaw et
  al.]{E. ~J. Bradshaw$^{1}$\thanks{Email: emma.bradshaw@nottingham.ac.uk},
  O. Almaini$^{1}$, W.~G. Hartley$^{1}$, K. T. Smith$^{2}$, C.~J. Conselice$^{1}$,
\newauthor   J.~S. Dunlop$^{3}$, C. Simpson$^{4}$,
  R.~W. Chuter$^{1}$, M. Cirasuolo$^{3}$, S. Foucaud$^{5}$,
\newauthor
  R.~J. McLure$^{3}$, A. Mortlock$^{1}$, H. Pearce$^{3}$\\$^{1}$School of Physics and
  Astronomy, University of Nottingham, University Park, Nottingham NG7
  2RD \\$^{2}$Royal Astronomy Society, Burlington House, Piccadilly, London W1J 0BQ \\$^{3}$Institute for Astronomy, University of Edinburgh, Royal
  Observatory, Edinburgh EH9 3HJ
  \\$^{4}$Astrophysics Research Institute, Liverpool John Moores
  University, Twelve Quays House, Egerton Wharf, Birkenhead CH41 1LD
  \\$^{5}$Department of Earth Sciences, National Taiwan Normal University
No.88, Sec. 4, Tingzhou Rd., Wenshan District, Taipei 11677, Taiwan }

\begin{document}

\date{Accepted 2013 April 24.  Received 2013 April 22; in original form 2012 November 7}

\pagerange{\pageref{firstpage}--\pageref{lastpage}} \pubyear{2012}

\maketitle

\label{firstpage}

\begin{abstract}
We investigate galactic-scale outflows in the redshift range 0.71 $\leq z\,\leq$ 1.63, using 413 K-band selected galaxies observed in the spectroscopic follow-up of the UKIDSS Ultra-Deep Survey (UDSz). The galaxies have an average stellar mass of $\sim$10$^{9.5}$~M$_{\odot}$ and span a wide range in rest-frame colours, representing typical star-forming galaxies at this epoch. We stack the spectra by various galaxy properties, including stellar mass, [O\,\textsc{ii}] equivalent width, star-formation rate, specific star-formation rate and rest-frame spectral indices. We find that outflows are present in virtually all spectral stacks, with velocities ranging from 100-1000 km\,s$^{-1}$, indicating that large-scale outflowing winds are a common property at these redshifts. The highest velocity outflows ($>$500 km\,s$^{-1}$) are found in galaxies with the highest stellar masses and the youngest stellar populations. Our findings suggest that high velocity galactic outflows are mostly driven by star-forming processes rather than AGN, with implied mass outflow rates comparable to the rates of star formation. Such behaviour is consistent with models required to reproduce the high-redshift mass-metallicity relation.
\end{abstract}

\begin{keywords}
galaxies: active -- quasars: general -- galaxies: evolution
\end{keywords}


\section{Introduction}\label{Introduction}
One of the key unanswered questions in extragalactic astronomy is the origin of the galaxy red sequence, and the strong bimodality in the colour--absolute-magnitude plane \citep[e.g.][]{{1977ApJ...216..214V}}. This bimodality appears to be in place from the local Universe up to redshifts of at least $z\, \sim 2$ \citep[e.g.][]{{2007MNRAS.380..585C}, {2008MNRAS.385.2049G}, {2011MNRAS.413.1678C}}.  More than half of all present-day early-type galaxies appear to have evolved into quiescent galaxies by $z\,\sim 1$ \citep[e.g.][]{{2005ApJ...634..861Y}}. From $z\, \sim 1$ to the present day, the red sequence almost doubles in mass \citep[e.g.][]{2006ApJ...651..120B}, which is thought to be due to galaxies quenching their star formation and therefore moving across the colour-absolute magnitude plane from the blue cloud to the red sequence, from which point on they evolve passively.

One mechanism for terminating star formation is via galactic-scale winds, which eject the gas from a galaxy and deplete the material needed to form stars. Once star formation has been terminated, the galaxy becomes a passive (quiescent) galaxy, at which point it migrates onto the red sequence. Galactic outflows can be driven by stellar processes (e.g. from stellar winds and supernovae) or from active galactic nuclei \citep[AGN; e.g.][]{{1998A&A...331L...1S},{2011MNRAS.415L...6K}} and may also play a key role in regulating the metal content of a galaxy and enriching the intergalactic medium (IGM).

In starburst galaxies, the energy provided by winds from young stars and supernovae is known to drive super-heated bubbles of metal enriched plasma \citep[e.g.][]{{1990ApJS...74..833H}}. Such bubbles expand, sweeping up the interstellar medium as it accelerates to velocities of a few hundred\,km\,s$^{-1}$. There is recent evidence \citep[e.g.][]{{2012ApJ...755L..26D}} to suggest that such winds can reach velocities of $>$1000\,km\,s$^{-1}$ if they have been driven by starburst galaxies with a high surface density of star formation.

In addition to starburst driven winds, feedback from AGN is now required in models of galaxy formation, to suppress the formation of the most massive galaxies \citep[e.g.][]{{2003ApJ...599...38B}}. A number of models use AGN feedback in the form of galactic-scale outflows, sufficient to remove the majority of gas, quench star formation and thereby create massive, passive galaxies \citep[e.g.][]{{2005Natur.433..604D},{2005ApJ...630..705H}}. Low-luminosity AGN activity may then suppress any further star formation as it `sweeps up' the remaining gas which would otherwise have been used to form more stars \citep[e.g.][]{2011arXiv1101.4230H}.  Alternative models use `radio-mode' feedback in the form of expanding bubbles within the hot intergalactic medium \citep[e.g.][]{2006MNRAS.370..645B}. The precise form of AGN feedback remains unclear, and a combination of radio-mode feedback and AGN-driven winds may be required to suppress star formation sufficiently \citep[]{2011MNRAS.415.2782V}.

Galactic winds can be detected through the study of many different emission and absorption lines, but in particular, spectra in the restframe UV are ideal for probing the interstellar medium (ISM). This part of the spectrum is rich in interstellar absorption lines, such as Mg\,\textsc{ii} (2796, 2803 \AA) and Fe\,\textsc{ii} (2344, 2374, 2382, 2586, 2600 \AA), which can be used to trace cool interstellar gas. Outflows with velocities of up to $\sim$600\,km\,s$^{-1}$ are often detected in starburst galaxies and appear to be present at all epochs \citep[][]{{2009ApJ...692..187W},{2010ApJ...719.1503R},{2011arXiv1104.0681C}} and are thought to be the product of supernova explosions. Outflow signatures can also be detected in the far UV with absorption lines such as C\,\textsc{iv}, Si\,\textsc{ii}, Si\,\textsc{iv} etc \citep{2004ApJ...604..534S}. Outflows with higher velocities are thought to be AGN-driven \citep[][]{{2006ApJ...653...86T},{2009ApJ...696..214S},{2011ApJ...733L..16S}}, although (as previously mentioned) there is recent evidence from \cite{2012ApJ...755L..26D} to suggest they could also be starburst driven.

\cite{2007ApJ...663L..77T} discovered relic outflows with velocities $>$1000\,km\,s$^{-1}$ in 10 out of 14 high resolution spectra of highly-luminous post-starburst galaxies at a redshift of $z\sim 0.6$ using the Mg\,\textsc{ii} $\lambda \lambda$ 2795.53, 2802.71 \AA\, doublet. Post-starburst (also referred to as E+A) galaxies are those in which star formation has recently been terminated and are assumed to be in the transition to becoming an early-type galaxy. The high occurrence of high-velocity outflows in bright, post-starburst galaxies suggests that these outflows could be the quenching mechanism responsible for terminating star formation. \cite{2007ApJ...663L..77T} suggest that these winds may be AGN-driven, given the high velocity of the outflows. It is unclear, however, if this is the mechanism by which more typical galaxies migrate to the red sequence.

Work on winds and galactic-scale outflows has also been carried out at higher redshifts ($z>2$). For example, \cite{2002ApJ...569..742P} studied galactic winds by observing a gravitationally lensed galaxy with a redshift $z=2.72$. They utilised UV absorption lines to measure outflows and found that although the bulk of the interstellar medium has a velocity $\sim$ 250\,km\,s$^{-1}$, there are also outflows present with velocities of at least three times this value. Pettini et al. deduced from this work that (if the outflow is spherical in geometry) the mass outflow rate is greater than the star-formation rate. However, they could not determine whether the gas stays gravitationally bound to the galaxy and hence whether the winds could terminate star formation. \cite{2003ApJ...588...65S} investigated the presence of outflows in Lyman break galaxies (LBGs) at $z \sim 3$ by comparing redshifts derived from interstellar lines with Lyman-alpha emission. They found outflows exist at this redshift with a wide range of velocities and the average offset between the velocities of the interstellar absorption lines and the Lyman-alpha emission was found to be $\sim$650\,km\,s$^{-1}$. Finally, \cite{2005MNRAS.359..401S} observed a SCUBA galaxy at $z=2.4$ and found a gas outflow with a velocity of 220\,km\,s$^{-1}$, which they interpret as a starburst-driven wind. Outflows at higher redshifts seem extremely common, but outflows in galaxies with an intermediate redshift have been studied very little until recently.

In this work we investigate galactic scale outflows up to redshifts $z \sim 1.6$, the epoch when the red sequence is still being rapidly assembled and AGN activity peaks.  We use a sample of $K$-band selected galaxies, probing more typical galaxies than the work by \cite{2007ApJ...663L..77T}. Our work differs from other recent studies \citep[][]{{2009ApJ...692..187W},{2010ApJ...719.1503R},{2011arXiv1104.0681C}} by the inclusion of many very red galaxies at $z>1$, including many  
on the galaxy red sequence. We use 413 galaxy spectra with redshifts in the range $0.71 < z < 1.63$, stacking the spectra by galaxy colours, stellar mass and spectral properties. Our aim is to determine the relationship between galaxies properties and the occurrence of high-velocity outflows, as measured from the Mg\,\textsc{ii} absorption feature.

Where relevant, we adopt a concordance cosmology in our analysis; $\Omega_M=0.3$, $\Omega_{\Lambda}=0.7$, $h={\rm H}_0/100 ~{\rm\,km\,s}^{-1}{\rm Mpc}^{-1} =0.7$ and $\sigma_8=0.9$. Section 2 describes the photometric and spectroscopic data used in this work, while Section 3 describes the methodology of our stacking and line-fitting analysis. Section 4 presents the results of our work based on an analysis of Mg\,\textsc{ii} feature, while Section 5 discusses the implications for terminating star formation. Section 6 presents a summary of our conclusions.

We separate the spectroscopic data used in this analysis into those obtained with the FORS2 instrument and those obtained with VIMOS, due to the differing resolutions of the instruments. We describe the VIMOS methodology in the main body of the text and discuss the details of the FORS2 analysis in the appendix.


\section{Sample Selection}\label{Data}

\subsection{UDS Photometric Data}

To investigate the prevalence and velocity structure of outflows, our galaxy sample is drawn from the Ultra Deep Survey (UDS). The UDS is the deepest of the five surveys that comprise the UKIRT Infrared Deep Sky Survey (UKIDSS; Lawrence et al. 2007). The UDS covers an area of 0.77 square degrees, centred on the Subaru-XMM Deep Field (SXDF), and is the deepest near-infrared survey over such a large area to date. In this work we use data from the UKIDSS data release 8 (DR8), reaching depths of $K= 24.6, H= 24.2, J= 24.9$ (AB, $5\sigma$, $2\arcsec$ aperture). In addition to near infrared data, the field is covered by deep optical data in the B, V, R, i$^{\prime}$ and z$^{\prime}$-bands from the Subaru-XMM Deep Survey, achieving depths of $B = 28.4, ~V = 27.8, ~R = 27.7, ~i^{\prime} =27.7$ and $z^{\prime} = 26.7$ \citep[AB, $3\sigma$, $2\arcsec$]{{2008ApJS..176....1F}}. {\it Spitzer} data reaching $5\sigma$ depths of 24.2 and 24.0 (AB) at $3.6\mu$m and $4.5\mu$m respectively is provided by the {\it Spitzer} UDS Legacy Program (SpUDS, PI:Dunlop) with U-band data taken with the Canada-France-Hawaii Telescope (CFHT) using the Megacam instrument ($U_{AB} = 25.5$; Foucaud et al. in prep). Although spectroscopy is used for the majority of the analysis, we also utilise UDS imaging to calculate galaxy stellar masses and to compare star-formation rates derived from emission lines with SED fitting.

\subsection{UDS Spectroscopic Data}

The spectra used in this analysis were taken from the ESO Large Programme 180.A-0776 (UDSz; PI: Almaini), using a combination of the VIMOS and FORS2 instruments (VIsible Multi-Object Spectrograph and FOcal Reducer and low dispersion Spectrograph respectively), installed on the ESO VLT.

The UDSz programme was designed to target a large $K$-band selected sample of $\sim$3500 galaxies across the UDS field, with photometric redshifts used to preferentially target galaxies at $z_{phot} > 1$. VIMOS was used to target optically brighter systems ($i_{AB}^{\prime} < 24$ or $V_{AB} < 25$) whilst FORS2 was used for fainter, redder systems ($i_{AB}^{\prime} < 25$ and $V_{AB} > 24$) with deliberate overlap in the selection criteria.

The UDSz survey used eight pointings of VIMOS in multi-object mode to sample an area of $0.5$ sq degrees across the UDS field, typically observing over 300 targets at each position. For each of these pointings we obtained 4.5 hours exposure with the LR Blue grism and 2.6 hours with the LR Red grism with resolving powers of $R= \lambda / \delta\lambda =180, 210 $ respectively, with the majority of targets observed with both grisms to maximise wavelength coverage. With the LR Red and LR Blue grisms we obtained an average RMS of $\sim$1.1~\AA\ and $\sim$0.93~\AA\ in the wavelength calibration respectively, and for the FORS2 data, this decreases to $\sim$0.24~\AA\ . There were 20 FORS2 fields coincident with the central part of each VIMOS tile; each of which received 5.5 hours exposure with the 300I grism, typically observing 35-40 faint galaxies per field with a resolving power of $R=660$. In total, 2881 objects were targeted with VIMOS and 802 with FORS2.

The VIMOS and FORS2 spectra were reduced separately at different institutions. For more information on the FORS2 reduction, see \cite{2012MNRAS.422.1425C}. The VIMOS spectra were firstly extracted and reduced using the VIMOS Interactive Pipeline Graphical Interface \citep[VIPGI;][]{{2005PASP..117.1284S}}. The zeroth order lines and known sky lines were then removed from the 1D spectra. The spectra were trimmed to remove regions badly affected by fringing and then VIMOS LR Red and LR Blue spectra were spliced together (where applicable) to yield typical spectra with a total wavelength range of 3700-9500 \AA. The spectra were resampled and spliced together using the mean S/N in the overlap region (6000-6600 \AA) to weight the spectra. The average continuum S/N measurement was found to be 6.46 \AA\,$^{-1}$ for the red spectra and 4.49 \AA\,$^{-1}$ in the blue. The spectra we use in this analysis are those with highly secure redshifts (see below), for which the S/N is typically higher. In order to obtain the high S/N required  for the detailed study of the Mg\,\textsc{ii} absorption line, however, we create composite spectra by stacking the individual galaxy spectra in the restframe (see Section 3.2).

\subsubsection{Spectroscopic Redshifts}\label{Sample}

Redshifts for the spectra were obtained using two packages: SGNAPS and EZ \citep{2008ASPC..394..239F}. EZ determines redshifts using the cross-correlation of templates, the solutions of which can then be checked manually. We used a selection of default templates and also our own templates which were constructed after a first pass of redshift determination. These templates will be presented in Bradshaw et al. (in preparation).  Overall, we obtained 1512 highly secure redshifts from the primary $K$-selected sample spanning a redshift range $0 < z < 4.8$, including several hundred in the `redshift desert' ($1.4 < z < 2.2$). A number of $z > 6$ candidates were also confirmed (see Curtis-Lake et al. 2012).

For the outflow analysis it was important to use only those spectra with a secure, accurately-determined redshift, and therefore we chose to use only spectra with strong [O\,\textsc{ii}] emission at 3727.5\AA\,, which is easily identifiable and close to Mg\,\textsc{ii} in wavelength. We also further limit the sample by selecting only those spectra with [O\,\textsc{ii}] of sufficient strength to yield a redshift accurate to $\sim$100\,km\,s$^{-1}$, as determined by fitting to the centroid of the emission line. An error of this magnitude is typically twice the error in the VIMOS wavelength calibration. By using [O\,\textsc{ii}] as a zero-point to define the systemic redshift we ensure that the position of Mg\,\textsc{ii} is measured consistently for all spectra. Although we cannot resolve the lines, [O\,\textsc{ii}] is actually a doublet with emission lines at 3726.1 \AA\, and 3728.8 \AA\,. We assume a doublet ratio of 8:9, as observed by \cite{2009ApJ...692..187W} using higher-resolution spectra for a sample of galaxies with similar redshifts and stellar masses. This is equivalent to an electron density of $\sim$5$\times$10$^{8}$\,m$^{-3}$. The centre of the [O\,\textsc{ii}] emission therefore lies at 3727.5 \AA\,. We use this wavelength throughout the rest of our analysis.

In order for both [O\,\textsc{ii}] and Mg\,\textsc{ii} to be visible in the observed spectra we effectively restrict our analysis to redshifts in the range $0.71<z<1.41$ for the VIMOS data, extending up to $z \simeq 1.63$ for the FORS2 data. Finally, the spectra identified as clear AGN were removed from this data analysis; this included those identified as X-ray AGN by \cite{2008ApJS..179..124U} and radio AGN by \cite{2006MNRAS.372..741S}, \citep[see][for more information]{{2011MNRAS.415.2626B}} as well as those spectra displaying emission lines commonly identified as originating from AGN, such as [C\,\textsc{iv}] and C\,\textsc{iii}] emission. Spectra with badly subtracted skylines were also excluded from this analysis, leaving a sample total of 413 galaxy spectra, with 341 from VIMOS and 72 from FORS2. The 413 galaxies in the sample have an average stellar mass of $\sim$10$^{9.6}$ M$_{\odot}$ and display a wide range of rest-frame colours (see Section 4). Given the primary selection in the $K$-band, our sample will be broadly representative of typical  galaxies at this epoch. The most passive galaxies (with little or no star formation) will be somewhat under-represented, due to the requirement for [O\,\textsc{ii}] emission in our sample selection, but we note that the FORS2 sample in particular contains many galaxies located on the galaxy red sequence (see Section~\ref{U-B}).

\subsubsection{Stellar Masses}\label{MassMethod}

The stellar masses and magnitudes used in this work are measured using a multi-colour stellar population fitting technique where we fit to the $UBVRizJHK$ bands and IRAC Channel 1 and 2 bands. A large grid of synthetic SEDs are constructed from the stellar population models of Bruzual \& Charlot (2003), assuming a Chabrier initial mass function. The star-formation history is characterised by an exponentially declining model with various ages, metallicities and dust extinctions. These models are parametrised by an age of the onset of star formation, and by an e-folding time such that
\begin{equation}
SFR(t) \sim SFR_{0}\times e ^{-\frac{t}{\tau}}.
\label{eq:SFRexp}
\end{equation}
where the values of $\tau$ can range between 0.01 and 13.7 Gyr, while the age of the onset of star formation ranges from 0.001 to 13.7 Gyr. The metallicity ranges from 0.0001 to 0.1, and the dust content is parametrised by $\tau_{v}$, the effective $V$-band optical depth for which we use values $\tau_{v}$ = 0, 0.2, 0.4, 0.6, 0.8, 1.0, 1.33, 1.66, 2, 2.5.

To fit the SEDs we first scale them to the $K$-band magnitude of the galaxy we are fitting to. We then fit each scaled model template in the grid of SEDs to the measured photometry of the galaxy. We compute the chi-squared values for each template and select the best fit. From this we obtain a best-fit stellar mass and best-fit absolute magnitudes. Further  details of the stellar mass determination can be found in \cite{2013MNRAS.tmp.1080H} and Mortlock et al. (submitted).


\section{Creating Composite Spectra}

\begin{figure*}
\begin{center}
\hspace*{-20pt}
\includegraphics[width=18.0cm]{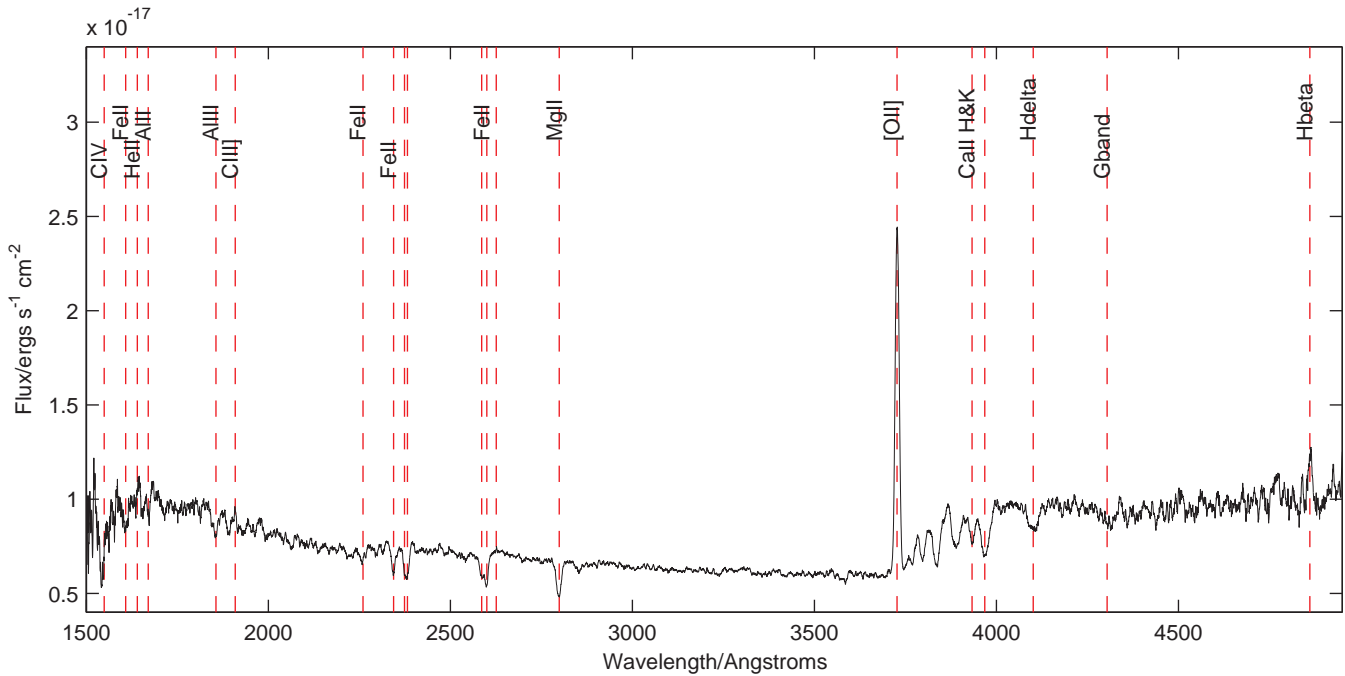} 
\end{center}
\caption{\small{The co-added stack of the  341 VIMOS spectra used in this work, all selected to ensure that  [O\,\textsc{ii}] and Mg\,\textsc{ii} were within the observed spectral range. Other emission and absorption features are also shown. Flux is shown in ergs~s$^{-1}$~cm$^{-2}$.}}
\label{fig:all_stacked_413}
\end{figure*}

We use the Mg\,\textsc{ii} absorption feature as a tracer of outflowing galactic winds because it is a sensitive tracer of cool gas in the ISM. To study the properties of the Mg\,\textsc{ii} $\lambda$$\lambda$ 2796, 2803 doublet, we performed a stacking analysis with our spectra, as there are only a handful of galaxies which are bright enough to individually measure the structure of the Mg\,\textsc{ii} profile. Higher S/N composites allow us to study the relationship between the average outflow properties and a range of galaxy characteristics.

The stacking analysis was conducted for the VIMOS and FORS2 data separately as the resolutions of the two instruments are different. For clarity, we place details of the FORS2 analysis in the appendix and describe here the main analysis with the VIMOS spectra, as these data comprise the bulk of the sample (341 VIMOS spectra compared to 72 FORS2 spectra).

\subsection{Co-addition of UDSz Spectra}\label{stacking_method}

The redshifts obtained during the data reduction process were used to shift the spectra to the restframe whilst oversampling the spectra onto a finer dispersion axis so as not to lose information. In order to confirm the reliability of the zero-point redshift derived from the [O\,\textsc{ii}] line at 3727.5 \AA\,, we calculate the offset between [O\,\textsc{ii}] and the CaII K absorption line at 3933.7 \AA\,, which should be unaffected by any systemic redshift offset from outflowing gas, as it is primarily a stellar absorption line.  From the composite spectrum of the VIMOS sample we find that the centroid of the Ca\,\textsc{ii} absorption line is offset by 0.2 \AA\ ($\sim$ 20\,km\,s$^{-1}$) from the [O\,\textsc{ii}] emission, which is well below the value of our wavelength calibration error. We also compare the velocity offset between [O\,\textsc{ii}] and C\,\textsc{iii}], as this feature is towards the bluest end of our spectra and hence provides a good test of our calibration uncertainties over a wide wavelength range. The C\,\textsc{iii}] $\lambda$ 1909 \AA\, emission is known to be a doublet ($\lambda \lambda$ 1906.1, 1908.7 \AA\,), where the flux density of the lines depend particularly on the electron density of the emitting gas \citep{1992ApJ...389..443K}. With our data, we cannot resolve the C\,\textsc{iii}] lines, so we use a ratio of 2:3 with a centre at 1907.7~\AA\,, corresponding to an electron density of $\sim$5$\times$10$^{8}$\,m$^{-3}$ within star-forming regions, consistent with the electron density we implicitly assume for the [OII] doublet ratio \citep[e.g.][]{{2012A&A...539A..61T}}. The C\,\textsc{iii}] emission, was found to have an offset of 0.3 \AA\ ($\sim$ 30\,km\,s$^{-1}$) from [O\,\textsc{ii}], which is again below the error in our wavelength calibration. Similar values were obtained using the other composite spectra used in this work. These checks suggest that the [O\,\textsc{ii}] line provides a reasonable measure of the systemic redshifts.

Two subtly different routines were used to construct the composite spectra. In both methods, the continuum levels of all spectra were scaled by fitting the continuum in the restframe wavelength range 2850-3050 \AA\,, a featureless window near Mg\,\textsc{ii} (2800 \AA\,). The spectra were co-added, firstly by taking the mean pixel value at each wavelength increment (thus ensuring a high S/N) and secondly by using the median pixel value at each wavelength increment (thus ensuring the signal from any unusual galaxies do not dominate the co-added spectrum). The composite spectrum of all 341 VIMOS galaxies in our sample (using the median  method) is shown in Figure~\ref{fig:all_stacked_413}.  We find that the chosen method of stacking has no significant impact on the results presented, so we use the median stacking method throughout this work.

\begin{figure}
\begin{center}
\hspace*{-20pt}
\includegraphics[width=8.0cm]{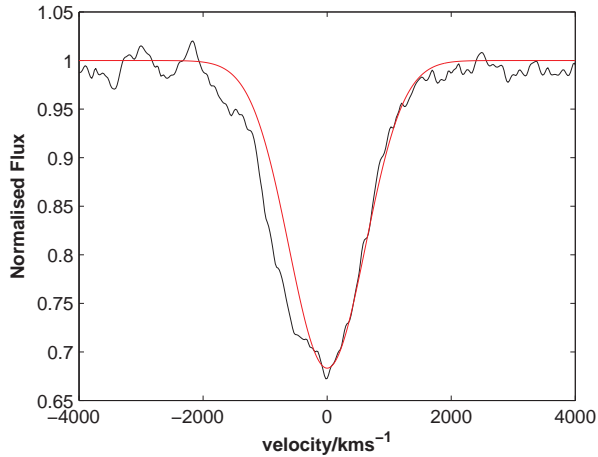} 
\end{center}
\caption{\small{The Mg\,\textsc{ii} absorption profile from a co-added spectrum containing all 341 VIMOS galaxies in our sample. The red curve shows the best-fitting Gaussian curve, centred at 0~km\,s$^{-1}$. Clear absorption blueward of the line is visible. By fitting a two-component model (see Section 3.2.1), we obtain a velocity of $\sim$340\,km\,s$^{-1}$ for the outflowing component.}}
\label{fig:MgII_all_stacked}
\end{figure}

A close-up of the Mg\,\textsc{ii} absorption profile from the stack of all 341 VIMOS galaxy spectra is shown in Figure~\ref{fig:MgII_all_stacked}. By fitting a Gaussian curve to the profile with its centroid at 0~km\,s$^{-1}$ we find clear evidence for excess absorption blueward of the fitted line, suggesting a component of outflowing gas. By fitting a Gaussian to the whole Mg\,\textsc{ii} profile (without fixing the centroid at 0\,km\,s$^{-1}$), we measure a blueshift of $\sim$1.5\AA\, which corresponds to an outflow velocity of $\sim$150\,km\,s$^{-1}$.  However, this simple fit is likely to underestimate the velocity of any outflowing component if there is a significant component of non-outflowing gas in the stack. We therefore use a method which fits two Gaussian profiles (as detailed in Section 3.2) to parameterise the intrinsic and outflowing components of Mg\,\textsc{ii}. This method yields an average wind velocity of 340\,km\,s$^{-1}$ for the outflowing component.

\subsection{Isolating Outflowing MgII}

In this Section we describe the decomposition of the Mg\,\textsc{ii} absorption profile into two separate components; a non-outflowing component (at the systemic velocity) and an outflowing component, with the velocity determined from a fitting procedure.

We assume that, in a composite spectrum of galaxies, there will be a sufficiently large number of spectra such that the absorption due to non-outflowing galactic gas will have a mean velocity of zero, matching the systemic velocity determined from the [O\,\textsc{ii}] line.  On the other hand, the outflowing gas from a galaxy will always have a zero or negative velocity, depending on the geometry of the outflow. We would not be able to see the positive velocity component of the outflow (should the outflow be symmetrical) due to the lack of continuum light sources along the line of sight.

It should be noted that the Mg\,\textsc{ii} equivalent widths that we derive below from the composite spectra do not necessarily provide a means for determining the column density of the absorbing gas. Although it is not obvious from our low-resolution spectra, the Mg\,\textsc{ii} absorption in an individual galaxy is typically very close to saturated (Weiner et al. 2009). Therefore, in this work, the absorption depths and measured equivalent width provide only an approximate indicator of the fraction of galaxies at a given velocity in the composite spectrum.

\subsubsection{Methodology: Extracting the Intrinsic and Outflowing MgII Components}\label{excess}

Firstly, to prepare the spectra for fitting, they were trimmed to a wavelength region that includes Mg\,\textsc{ii} and [O\,\textsc{ii}] (2550-3800 \AA ). The continuum in this region was fitted with a spline function and removed by dividing the spectrum by the fit.

To isolate and examine the intrinsic and outflowing gas, we complete an analysis which involves a careful fitting of two model Mg\,\textsc{ii} Gaussian absorption components; one representing the intrinsic (non-outflowing) component of Mg\,\textsc{ii} (at zero
velocity) and the other representing the outflowing component (at negative velocities). For the intrinsic Mg\,\textsc{ii}, we centre a Gaussian curve at 0\,km\,s$^{-1}$, allowing only the line depth to vary. A similar method was used by \cite{2009ApJ...692..187W} to successfully isolate the systemic component from higher resolution spectra. The depth and velocity of the outflowing component were free parameters.

Both Gaussians were restricted to have the same FWHM, to match the measured instrumental resolution of VIMOS. This width was determined using the [O\,\textsc{ii}] emission line in the composite spectra at 3727 \AA, allowing for the doublet nature of this feature.  As an alternative method we also determined the FWHM by allowing a free fit to the red side of the absorption profile. Both methods gave consistent results, as noted below.

The two components are fitted jointly (using a least-squares fit) to the absorption profile of Mg\,\textsc{ii}. The results of this analysis can be seen (for example) in the right-hand column of Figure~\ref{fig:OII_trial}, where the red curve represents the intrinsic Mg\,\textsc{ii}, the green curve represents the outflowing Mg\,\textsc{ii}, and the blue curve shows the overall fit to the absorption profile.

In addition, to test these results, we also created a more complex model of the absorption profile, modelling the Mg\,\textsc{ii} doublet with two components centred at 2796 \AA\, and 2803 \AA\,, each with $\sigma =$ 60\,km\,s$^{-1}$ and using a doublet ratio of 1.1:1, as observed in higher resolution spectra of galaxies at $z\sim 1$ by Weiner et al. (2009). This in turn was convolved with another Gaussian to model the response function of the instrument. The resulting model was then fitted to the absorption profiles with no noticeable difference in results between this and the two-Gaussian method previously described.

In order to test the robustness of this method further, we also determined the instrumental FWHM by first fitting a Gaussian profile to the red side of Mg\,\textsc{ii}, but otherwise following the same procedure as detailed above. This was in order to test that the FWHM of [O\,\textsc{ii}] was a fair representation of the instrumental resolution (after accounting for its doublet nature). This alternative Mg\,\textsc{ii} profile was then fitted to the data in Section 4. No significant differences were found between the two methods. The maximum variation of the outflowing velocity component from the method using a fixed FWHM is 110\,km\,s$^{-1}$ and the average deviation between the two methods is 60\,km\,s$^{-1}$, where the Mg\,\textsc{ii} with the free-fitting FWHM extracts higher velocity outflows (in the vast majority of cases). In every case, the two methods are both consistent with each other within the margin of error calculated using the method in Section~\ref{errors}. For the remainder of this analysis we therefore present results determined by fixing the FWHM at the instrumental value determined from the [O\,\textsc{ii}] emission line, as described above.

\subsubsection{Emission Filling and Resonant Scattering}\label{EmissionFilling}

In the work by \cite{2009ApJ...692..187W}, it was noticed that there was an emission feature present in the spectra of low-mass galaxies with bluer UV slopes on the red side of Mg\,\textsc{ii}. The emission was attributed to possible AGN activity and all galaxies showing signs of this feature were removed from subsequent analysis. More recent papers on the topic have discovered that the emission may be caused by resonance-line scattering from receding gas on the far side of the galaxies that host outflows, resulting in a P-Cygni-like line profile with redshifted emission and blueshifted absorption \citep[e.g.][]{{2011ApJ...734...24P},{2011ApJ...728...55R},{2012ApJ...759...26E},{2012ApJ...758..135K}}. This phenomenon can lead to significant emission underlying the absorption, which can in turn make the outflow velocity hard to determine. Here we investigate the effect that emission filling has on our ability to extract an accurate outflow velocity from our data.

We obtain a general insight as to the effects of emission filling by simulating artificial spectra of Mg\,\textsc{ii} with a P-Cygni-like profile whilst varying both the amount of emission and the outflow velocity. We then attempt to re-extract the outflow velocity using the Mg\,\textsc{ii} model used in our analysis (as described in Section~\ref{excess}). We begin by modelling the P-Cygni profile shown in Figure 1 of \cite{2011ApJ...728...55R} with four overlapping Gaussians; two representing emission and two representing absorption (due to the Mg\,\textsc{ii} doublet). The profile in this paper shows extremely strong emission at a ratio of nearly 1:1 between the emission trough and the absorption peak. We then convolve this model with a Gaussian to match the instrumental resolution of our VIMOS spectra, allowing for the intrinsic resolution of the spectrum in \cite{2011ApJ...728...55R}. The strength of the emission components are progressively reduced until the spectrum consists of a pure absorption profile, and we also vary the velocity of the outflow from 100\,km\,s$^{-1}$ to 1000\,km\,s$^{-1}$.

\begin{figure}
\begin{center}
\hspace*{-20pt}
\includegraphics[width=8.0cm]{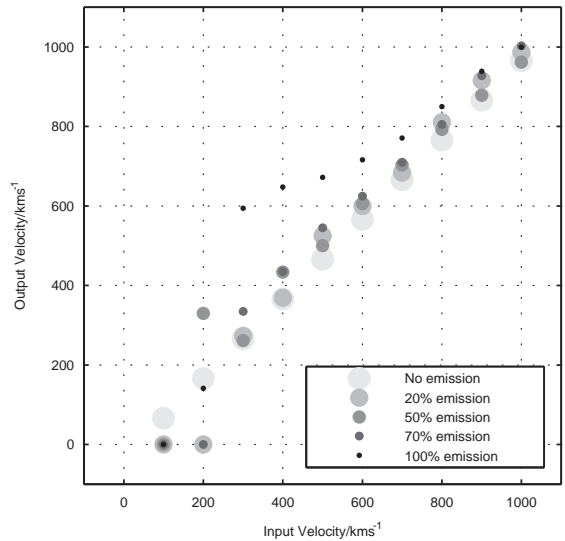} 
\end{center}
\caption{\small{This figure shows the results of simulations to quantify the impact of a P-Cygni-like emission feature on the Mg\,\textsc{ii} profile, which may affect our ability to determine an accurate outflow velocity. A range of outflow velocities and emission strengths are modelled. The percentages refer to the strength of the emission compared to the absorption. We find that outflow velocities can be re-extracted for low emission strengths, but velocities are systematically boosted upwards when the emission feature is very strong.}}
\label{fig:inputoutput}
\end{figure}

Figure~\ref{fig:inputoutput} shows the velocities that we re-extract from the artificial spectra for increasing absorption/emission ratios and increasing outflow velocities. For spectra in which the emission is 20$\%$ the strength of the absorption (or lower), we can easily extract the same outflow velocities that are input to the model. For spectra with an emission strength varying between 20$\%$ and 70$\%$ of the absorption, we must utilise caution in our analysis of any velocities lower than $\sim$300\,km\,s$^{-1}$. Finally, for emission which is the same strength as the absorption, we cannot accurately re-extract any velocity below 800\,km\,s$^{-1}$.

Inspection of our fits suggest that the outflow velocity we extract relies heavily on the blue side of the Mg\,\textsc{ii} profile, and hence the results are largely unaffected by emission filling at high outflow velocities (unless the emission component is extreme). At velocities below 300\,km\,s$^{-1}$, however, the outflows we infer may, in principle, be artificially boosted to higher values.

Finally, we compared the results of our simulated spectra with the real spectral stacks to determine if a strong emission component would be visible even at the low resolution of our VIMOS spectra, allowing for the noise in the data. We concluded that a strong Mg\,\textsc{ii} emission component ($>35\%$ of the absorption strength) would always be clearly visible given our signal to noise. The clearest evidence for a strong emission component was found in the low-mass stack presented in the top panel of Figure~\ref{fig:blue_mass}. Comparison with our model spectra suggest that the emission feature has an equivalent width comparable to $\sim$65$\%$ of the absorption component. Based on Figure~\ref{fig:inputoutput}, therefore, we believe the outflow velocity can be extracted down to an approximate limit of $\sim$300\,km\,s$^{-1}$, which is similar to the outflow velocity we extract for this spectrum. As our other composite spectra show little or no signs of emission filling, we determine that our method is generally sufficient for the rest of our analysis.

\subsection{Errors}\label{errors}

In order to obtain an estimate of the error involved in extracting a measurement for the equivalent width and velocity of the outflowing gas, we undertake a bootstrap analysis of our data. For every composite spectrum, we make an additional 100 spectra from sub-samples which are created by choosing a random selection of spectra within the same subset, but with replacement. All spectra are then resampled onto the same wavelength axis and the standard deviation on each pixel is calculated. The fitting procedure from Section~\ref{excess} is then repeated for each of the bootstrapped spectra, and we take the error of the velocity and column density to be one standard deviation from the mean of the 100 bootstrapped samples.

\subsection{Single-Gaussian Fit}\label{single_Gaussian}

In addition to the double-Gaussian fits described above, throughout this paper we also present the results of single-Gaussian fits, fixed at the systemic redshift and fitting only to the red side of the line. These fits allow the velocity and extent of any outflowing component to be visualised from the residuals blueward of the fitted line.  We follow the procedure in Section~\ref{excess} to fit a Gaussian profile which is centred at 0~km\,s$^{-1}$ using the instrumental FWHM determined from the [O\,\textsc{ii}] emission line. The depth of the profile is allowed to vary. The single-Gaussian fits help to illustrate whether an additional blue-shifted component is required.

\begin{figure*}
\begin{center}
\hspace*{-20pt}
\includegraphics[width=15.0cm]{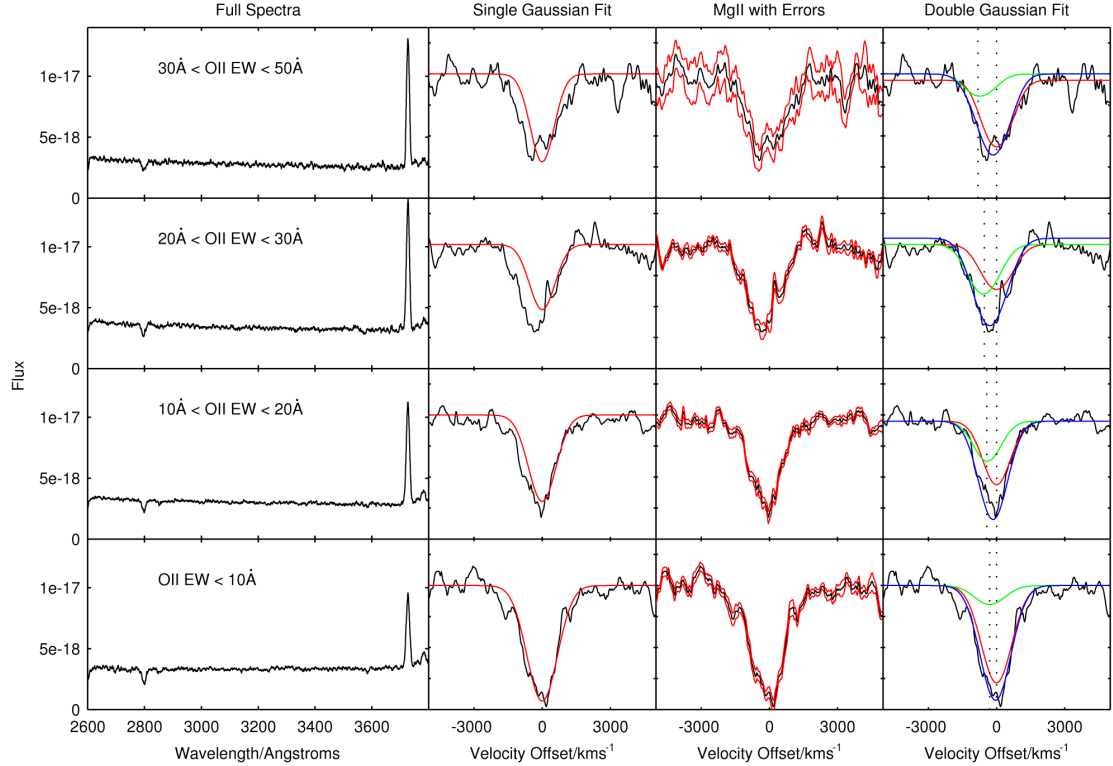} 
\end{center}
\caption{\small{The left panels show stacked spectra, arranged by decreasing [O\,\textsc{ii}] equivalent width, focussing on the wavelength range 2600-3800 \AA\ . The central-left panels show a close-up of the Mg\,\textsc{ii} absorption feature for each stack, along with the best-fitting single Gaussian (in red), obtained by fitting to the red side of the line only and fixed at zero velocity. The central-right panel shows the errors associated with each absorption profile, which were calculated using a bootstrap method. The right-hand panels show the breakdown of the absorption line fits used to calculate the velocity of the outflowing Mg\,\textsc{ii}, based on a two-component model. The red curve is the non-outflowing component, the green curve is the outflowing component, and the blue line represents the best-fit to the Mg\,\textsc{ii} line (the sum of the stationary and outflowing components). Flux is shown in ergs~s$^{-1}$~cm$^{-2}$.}}
\label{fig:OII_trial}
\end{figure*}

\begin{figure*}
\begin{center}
\hspace*{-20pt}
\includegraphics[width=15.0cm]{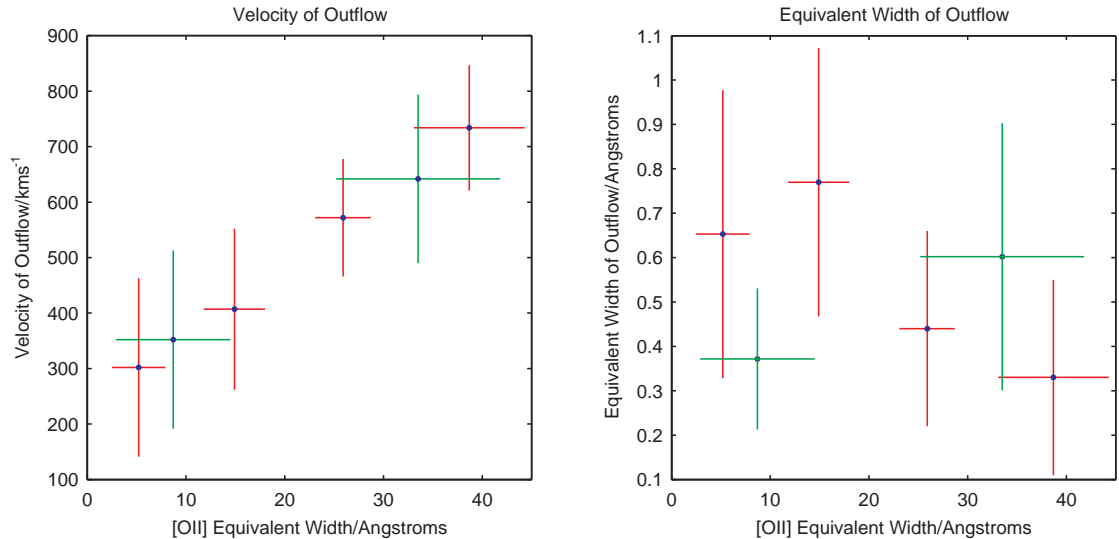} 
\end{center}
\caption{\small{The velocity (left panel) and the equivalent width (right panel) obtained for the outflowing component of Mg\,\textsc{ii} from the two-component fits to the stacked spectra shown in Figure 3, displayed as a function of [O\,\textsc{ii}] EW. The green points are based on the FORS2 data and the red points are based on the VIMOS sample. Data are plotted at the mean value of the [O\,\textsc{ii}] EW within a stack.}}
\label{fig:OII_EW_FORS2}
\end{figure*}


\section{Outflows in Galaxies Stacked by Property}\label{spectra}

In this section, we discuss the properties used to categorise the UDSz galaxies into stacks and the results of our subsequent fitting analysis. This section contains results from both the VIMOS sample and the FORS2 sample, although the detailed methodology for analysing the FORS2 spectra is contained in the appendix.

\subsection{Star Formation}\label{OII_EW}

The equivalent width of [O\,\textsc{ii}] emission at 3727\AA\, has long been recognised as an approximate indicator for the strength of star formation \citep[e.g.][]{{2002MNRAS.332..283R},{2004AJ....127.2002K}}. In the absence of significant dust or AGN activity, strong [O\,\textsc{ii}] emission indicates a significant burst of recent star formation. Therefore, in order to investigate the relationship between star formation activity and Mg\,\textsc{ii} outflows, we measured the equivalent width of [O\,\textsc{ii}] emission at 3727\AA\, for all VIMOS spectra. The spectra were then divided into four sub-samples; those showing the most active star formation (EW$=$30-50~\AA\,), two intermediate samples (EW$=$20-30~\AA\,) and EW$=$10-20~\AA\,) and finally those with weak star formation (EW$<$ 10~\AA\,). These bins were chosen to ensure a sufficient number of spectra in each stack (in this case there are 37, 51, 84 and 88 spectra in each stack respectively) whilst encompassing a large range of equivalent widths. The co-added spectra can be seen in in the left-hand panels of Figure~\ref{fig:OII_trial}. A single-Gaussian fit was then performed on the Mg\,\textsc{ii} profile in each stack, centred at 0~km\,s$^{-1}$ (representing the non-outflowing component of the Mg\,\textsc{ii}), where the curve was fitted only to the red side of the Mg\,\textsc{ii} absorption line (see Section~\ref{single_Gaussian}). The excess Mg\,\textsc{ii} and hence any outflowing material can then clearly be seen from the absorption blueward of this profile.

In order to illustrate the errors  involved in the fitting process, the central-right-hand panels of Figure~\ref{fig:OII_trial} show the errors on the stacked spectra through a bootstrap analysis, which was detailed in Section~\ref{errors}. The errors are noticeably larger in the stacks containing fewer spectra.

The velocity and equivalent widths of the intrinsic and outflowing components of Mg\,\textsc{ii} in each sub-sample are then derived from the method in Section~\ref{excess}, using two-Gaussian fits, and are shown in the right-hand panel. The equivalent width and the velocities of the Gaussians representing the outflowing component were extracted and plotted in Figure~\ref{fig:OII_EW_FORS2} in red, along with equivalent points extracted separately from the FORS2 data (in green). The errors are calculated from the methodology in Section~\ref{errors}. Those spectra with a high [O\,\textsc{ii}] equivalent width clearly show the highest velocity Mg\,\textsc{ii} outflows, but we find no correlation between [O\,\textsc{ii}] equivalent width and the equivalent width of the outflowing component of Mg\,\textsc{ii}.

\subsection{M$_{B}$ Magnitude}

\begin{figure}
\begin{center}
\hspace*{-20pt}
\includegraphics[width=8.0cm]{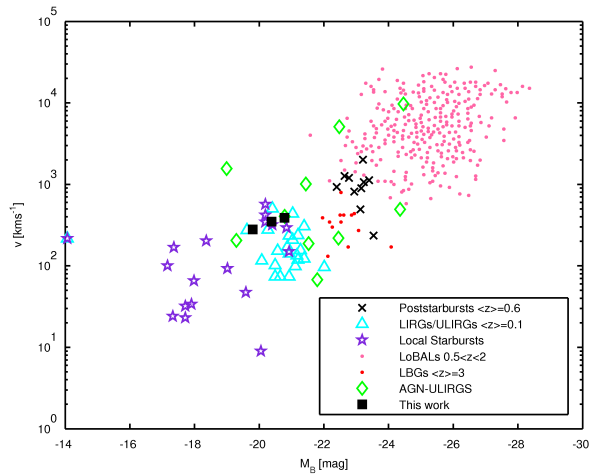} 
\end{center}
\caption{\small{This figure, adapted from Tremonti et al. (2007), shows the relationship between the velocity offset of low ionisation absorption lines and the B-band absolute magnitude. Our stacked data are represented by the black squares and correspond to the top end of the velocity offsets found in local galaxies classified as starbursts.}}
\label{fig:Tremonti}
\end{figure}

We divide the spectra into four sub-samples dependant on their absolute B-band magnitude; a bright sub-sample with $M_{B} < -21$, two intermediate sub-samples $-21 \leq M_B < -20.6$ and $-20.6 \leq M_B < -20.2$, and those spectra with $M_{B} \geq -20.2$. Each of these four samples were then co-added (see methodology in Section~\ref{stacking_method}), and curves were fitted in order to extract the intrinsic and outflowing Mg\,\textsc{ii} (see Section 3.2.1). For the sub-sample with the brightest B-band magnitude, an outflow velocity of 475 $\pm$ 114\,km\,s$^{-1}$ was extracted. Wind velocities of 350 $\pm$ 73\,km\,s$^{-1}$ and 389 $\pm$ 23\,km\,s$^{-1}$ were obtained for the two intermediate samples. We do not extract a velocity for the faintest sub-sample, as the spectrum is too noisy to fit to. We plot these results on a graph adapted from \cite{2007ApJ...663L..77T}, which compares the velocity shift to the average B-band magnitude of the galaxies (Figure~\ref{fig:Tremonti}). Our findings compare well with  the findings of previous authors, with outflow velocities at the upper end compared to local starburst galaxies and ultra luminous infrared galaxies (ULIRGS).

\subsection{Stellar Mass}\label{stellar_mass}

In this section we investigate the claim made by \cite{2009ApJ...692..187W} for a positive relation between galaxy stellar mass and outflow velocity. In order to match our sample to their work, we first limit our sample to blue galaxies with restframe $(U-B) <0.5$ to broadly match the DEEP2 sample from \cite{2009ApJ...692..187W}. We cannot perform a similar analysis on the FORS2 data, as there are only 27 spectra with a $(U-B)<0.5$, which are insufficient for a robust fit.

\begin{figure*}
\begin{center}
\hspace*{-20pt}
\includegraphics[width=18.0cm]{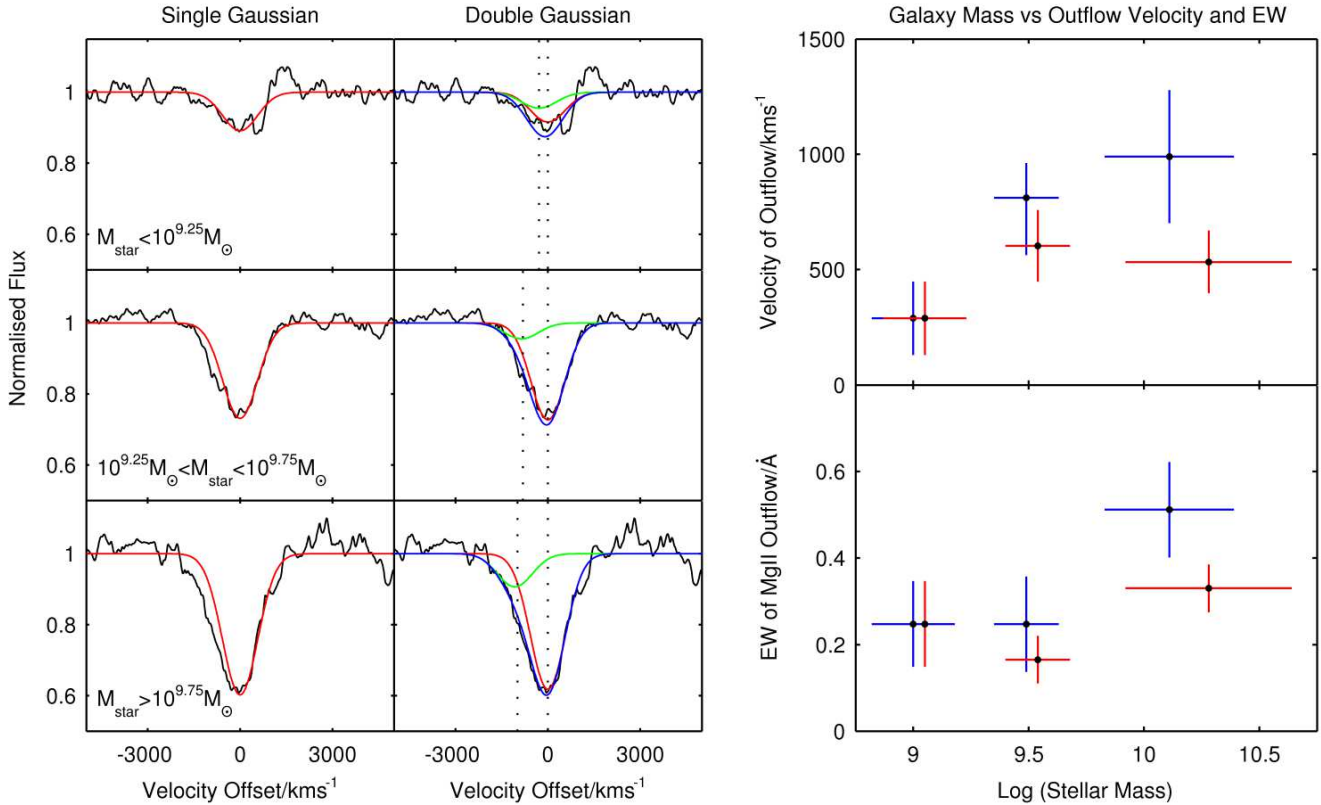} 
\end{center}
\caption{\small{The left-hand panels show a close-up of the Mg\,\textsc{ii} absorption feature for three spectral composites arranged by stellar mass, along with the best-fitting single Gaussian (in red), which represents the restframe Mg\,\textsc{ii}. The central panels show the breakdown of the absorption line fits used to calculate the velocity of the outflowing Mg\,\textsc{ii}. The red curve is the non-outflowing component, the green curve is the outflowing component, and the blue curve represents the best-fit to the Mg\,\textsc{ii} line (the sum of the stationary and outflowing components). These composite spectra are based on relatively blue galaxies with $(U-B) <0.5$ to allow a comparison with similar analyses in the literature (see Section 4.3). The right panel shows the equivalent width and velocity of the outflowing components as a function of stellar mass. In red we plot the complete VIMOS sample and in blue we plot blue galaxies with $(U-B) <0.5$ only. The mean stellar mass in the lowest-mass and medium-mass bins are the same, therefore for clarity we offset the points from each other.}}
\label{fig:blue_mass}
\end{figure*}

We divide our sample into three sub-samples, those with a stellar mass below 10$^{9.25}$M$_{\odot}$, those with a mass higher than 10$^{9.75}$M$_{\odot}$, and those with a mass between the two. We then create three composite spectra from these sub-samples and follow the same methodology as before to measure the velocity and equivalent width of outflowing material. As illustrated in Figure~\ref{fig:blue_mass}, we find evidence for higher velocity outflows in the galaxies with a higher mass. This trend is broadly consistent with \cite{2009ApJ...692..187W}, despite the different methods used to extract the velocities of the outflows.

We then put the red galaxies back into the sample and co-add the galaxies in the same three sub-samples as above. We follow the same methodology used throughout this paper to derive the velocity of the outflowing gas. The results are then also plotted on Figure~\ref{fig:blue_mass} in red. There are no red galaxies in the lowest mass sample and therefore the two points representing the co-added spectrum with the lowest mass are the same. We see a significant drop in the velocity of the outflow when we include the red galaxies, which suggests it could be colour and star-formation which most strongly determines the wind velocity.

As mentioned in Section~\ref{EmissionFilling}, we are cautious about extracting velocities using our two-component method for those spectra with signs of Mg\,\textsc{ii} emission filling. This is a common phenomenon for galaxies of lower mass \citep[e.g.][]{{2009ApJ...692..187W},{2012ApJ...759...26E}} and we do see emission in Figure~\ref{fig:blue_mass} slightly redward of the Mg\,\textsc{ii} absorption in our composite spectrum of low mass galaxies. As previously discussed, we model the Mg\,\textsc{ii} emission and absorption to find that the emission is $\sim$65$\%$ the strength of the absorption component and therefore we must use caution in the interpretation of results if the extracted velocities are below $\sim$300\,km\,s$^{-1}$ (Figure~\ref{fig:inputoutput}). In this case, the extracted velocities from the spectra composed from higher mass galaxies are above 700\,km\,s$^{-1}$, which suggests that there is a definite trend with mass, even if the extracted velocity of the lowest mass composite is potentially inaccurate.

\subsection{Star-formation Rate}\label{SFR}

Here we use [O\,\textsc{ii}] luminosity as a proxy for star-formation rate to determine whether there is a relationship with outflow velocity. We calculate the star-formation rate using the following equation from \cite{1989AJ.....97..700G}:
\begin{equation}
\mathrm{SFR}(\mathrm{[O\,\textsc{ii}]})(M_{\odot}\mathrm{yr}^{-1}) = 1.4 \times 10^{-41} L_{\mathrm{[OII]}}(\mathrm{ergs}\,\ \mathrm{s}^{-1}).
\end{equation}
We correct for dust extinction at intermediate redshifts following the methods derived in \cite{2012ApJ...746..124M}. The methods in this work take into account the difference in star-formation rate between starburst and passive galaxies and adjusts the reddening correction accordingly. We then compare our SFRs with values derived from SED
fitting (see Section~\ref{MassMethod}). We find generally good agreement. A straight-line fit to the relation between the two estimates of SFR gives a slope of 0.93, where SFRs derived from the [O\,\textsc{ii}] luminosity are slightly higher than those derived from SED fitting.

The [O\,\textsc{ii}] luminosity can also be affected by the metallicity and gas ionisation of the host galaxy, in addition to reddening. [O\,\textsc{ii}] is affected by the excitation state of the gas, which in turn is related to the ionisation \citep[e.g.][]{{2004AJ....127.2002K}}. We do not correct for galaxy metallicity or the ionisation parameter, as our spectra do not always include [O\,\textsc{iii}] or the H$\alpha$ and H$\beta$ lines that are commonly used to measure metallicity. In order to check that this did not affect the conclusions of this section, we also combine our spectra into bins dependent on star-formation rate derived from the SED fitting. Although the outflow velocities derived do vary slightly, the overall results of this section are not changed. We therefore continue with the SFRs we derive from [O\,\textsc{ii}] luminosity.

\begin{figure*}
\begin{center}
\hspace*{-20pt}
\includegraphics[width=15.0cm]{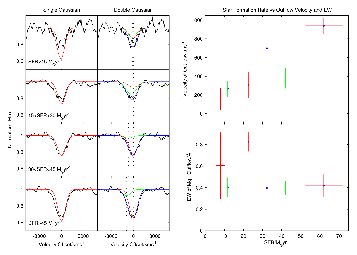}
\end{center}
\caption{\small{The left-hand panels show a close-up of the Mg\,\textsc{ii} absorption feature for each spectrum which have been co-added by star-formation rates, derived from [O\,\textsc{ii}] luminosity, along with the best-fitting single Gaussian (in red), which represents the restframe Mg\,\textsc{ii}. The central panels show the breakdown of the absorption line fits used to calculate the velocity of the outflowing Mg\,\textsc{ii}. The red Gaussian is the non-outflowing component, the green Gaussian is the outflowing component, and the blue line represents the best-fit to the Mg\,\textsc{ii} line (the sum of the stationary and outflowing Gaussian fits). The right panel (top) shows the equivalent width of the outflow as a function of star-formation rate and the right hand panel (bottom) shows outflow velocities as a function of star-formation rate.}}
\label{fig:SFR}
\end{figure*}

We divide our sample into four bins dependent on the derived star-formation rates, with the lowest bin having star-formation rates of $<$ 15 M$_{\odot}$yr$^{-1}$ and the highest bin having star-formation rates of $>$ 45 M$_{\odot}$yr$^{-1}$. We stack the spectra according to the method in Section 3.1 and fit a static and outflowing component to the Mg\,\textsc{ii} profile using the method in Section 3.2.1. We find that the highest velocity outflows are hosted by the galaxies with the highest star-formation rates. This can be seen in Figure~\ref{fig:SFR}. This is consistent with the results already obtained, that the outflow velocity increases with host galaxy mass and with [O\,\textsc{ii}] equivalent width (which is approximately proportional to specific star-formation rate) and suggests that outflow velocity depends on both parameters. It is also consistent with results from Weiner et al. (2009) who find tentative evidence for an increase in outflow velocity with star formation rate.

\subsection{Specific Star-formation Rate}\label{SSFR}

\begin{figure*}
\begin{center}
\hspace*{-20pt}
\includegraphics[width=15.0cm]{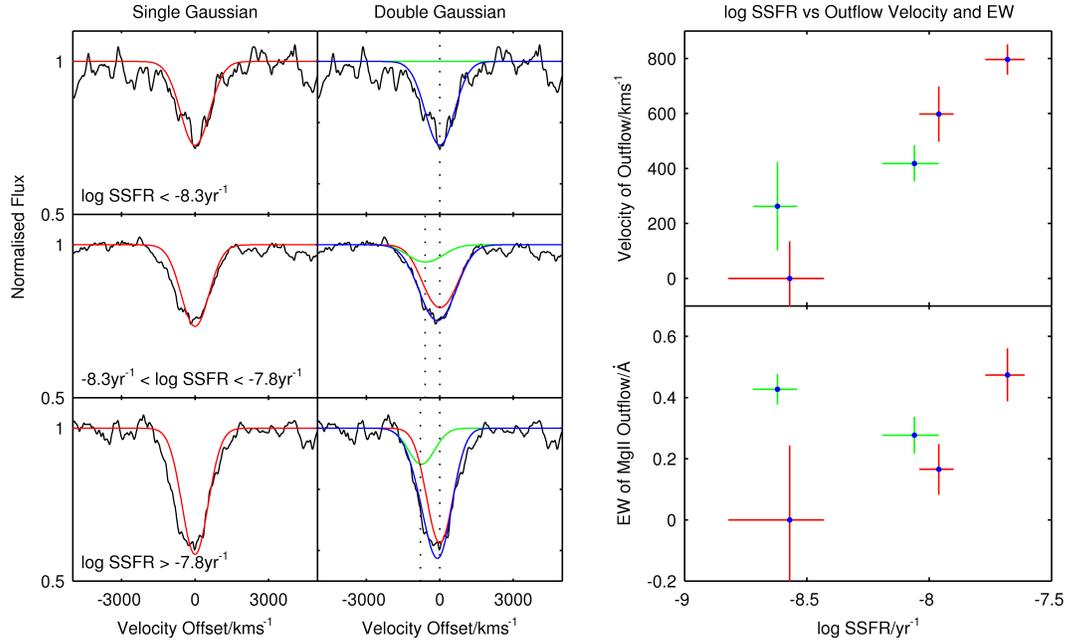}
\end{center}
\caption{\small{The left-hand panels show a close-up of the Mg\,\textsc{ii} absorption feature for each spectrum which have been co-added by specific star-formation rates, using the star-formation rates derived from [O\,\textsc{ii}] luminosity and galaxy masses, along with the best-fitting single Gaussian (in red), which represents the restframe Mg\,\textsc{ii}. The central panels show the breakdown of the absorption line fits used to calculate the velocity of the outflowing Mg\,\textsc{ii}. The red Gaussian is the non-outflowing component, the green Gaussian is the outflowing component, and the blue line represents the best-fit to the Mg\,\textsc{ii} line (the sum of the stationary and outflowing Gaussian fits). The right panel (top) shows the equivalent width of the outflow as a function of stellar mass and the right hand panel (bottom) shows outflow velocities as a function of specific star-formation rate.}}
\label{fig:SSFR}
\end{figure*}

We use the star-formation rates derived in Section~\ref{SFR} and the galaxy masses calculated in Section 2.2.2 to calculate the specific star-formation rate (SSFR) for the galaxy sample. We show the results in Figure~\ref{fig:SSFR}. A trend between excess Mg\,\textsc{ii} and SSFR is very apparent from the single Gaussian fits. From fitting two Gaussians, it is seen that the galaxies with the highest SSFRs have the highest velocity outflows (but again there is no trend with the equivalent width).

As expected, our results suggest a strong link between intense bursts of star formation and the occurrence of high velocity winds.

\subsection{Stellar Population}\label{definitions}

In this section, we use various spectral indices to determine galactic properties, and therefore divide our galaxies into sub-samples accordingly. First, we define the spectral indices that we will use to classify our spectra.

The first of these spectral indices is the $D(4000)$ index, which is a measure of the strength of the 4000\AA\ break:
\begin{equation}
D(4000) = \frac{F(4000-4100)}{F(3850-3950)}.
\end{equation}
The 4000\AA\ break arises from the lack of blue light from older stellar populations, the large accumulation of absorption lines from ionised metals, and the end of the Hydrogen Balmer series \citep[e.g.][]{{1999ApJ...527...54B},{2003MNRAS.341...33K}}.  The $D(4000)$ spectral index will be small for galaxies with young stellar populations and larger for older, metal-rich galaxies. The $D(4000)$ index was originally defined by \cite{1983ApJ...273..105B} as the ratio of the average flux density between the wavelengths 4050--4250 \AA\, and 3750--3950 \AA\,. The definition of $D(4000)$ from \cite{1999ApJ...527...54B} that we use in this work is defined with narrower bands in comparison and is therefore less sensitive to reddening effects.

\cite{2008A&A...482...21C} use the mean spectral flux between 2700-3100\AA\ and 3100-3500\AA\ (in the rest frame) to define the spectral index C(29-33):
\begin{equation}
C(29-33) = -2.5 \log\left(\frac{F(2900)}{F(3300)}\right).
\end{equation}
High values for the $C(29-33)$ index indicate fewer very young stars in the galaxy, with a galaxy classified as `passive' if it has an index $C(29-33)>0.5$ \citep{2008A&A...482...21C}. The individual galaxy spectra in the UDSz with the highest $C(29-33)$ indices show no (or very little) [O\,\textsc{ii}] emission, prominent metal absorption and a very red continuum.

\begin{figure}
\begin{center}
\hspace*{-20pt}
\includegraphics[width=8.0cm]{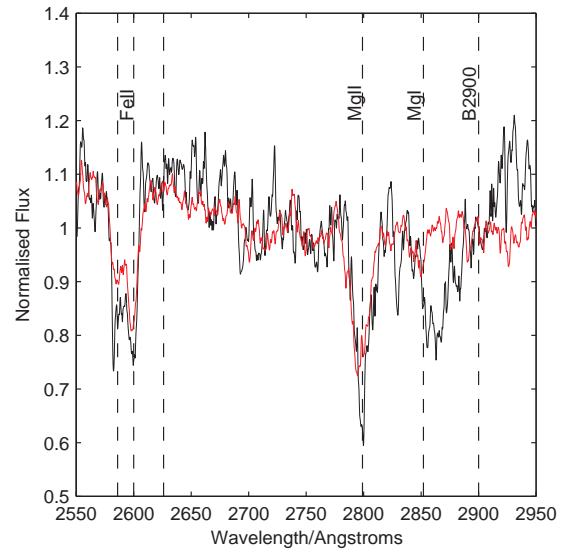} 
\end{center}
\caption{\small{A comparison between spectra with a low $B(2900)$ index (red) and the a high $B2900$ index (black). The difference in gradient of the continuum around $B2900$ is apparent, as well as the shape and strength of the Mg\,\textsc{ii} and Fe absorption lines.}}
\label{fig:B2900}
\end{figure}

\cite{1997ApJ...484..581S} use a break in galaxy spectra at 2900\AA\, to distinguish between galaxies with older  stellar populations:
\begin{equation}
B(2900) = \frac{F(2915-2945)}{F(2855-2885)}
\end{equation}
The $B(2900)$ index is used to estimate the age of the stellar population where higher values of this index indicate older stellar populations, measuring redder colours and stronger metal absorption. The $B(2900)$ index is defined over a narrow spectral range and therefore it is largely independent of reddening; it is determined by the opacity of the metals on the blue side of the spectral feature \citep{1997ApJ...484..581S}. A comparison between galaxies with a high and low $B(2900)$ index is shown in Figure~\ref{fig:B2900}.

The galaxies were divided into four sub-samples for each of the spectral indices (in the case of the VIMOS data) and co-added according to the methods described in Section 3.2.1. The sub-samples were chosen so that they span the available range of values for each spectral index, but with at least 50 spectra in each stack in order to maximise the S/N. Data from FORS2 were also used wherever possible (see appendix for details), extending the analysis towards older, redder galaxies beyond the reach of VIMOS at $z>1$.

\begin{figure*}
\begin{center}
\hspace*{-20pt}
\includegraphics[width=15.0cm]{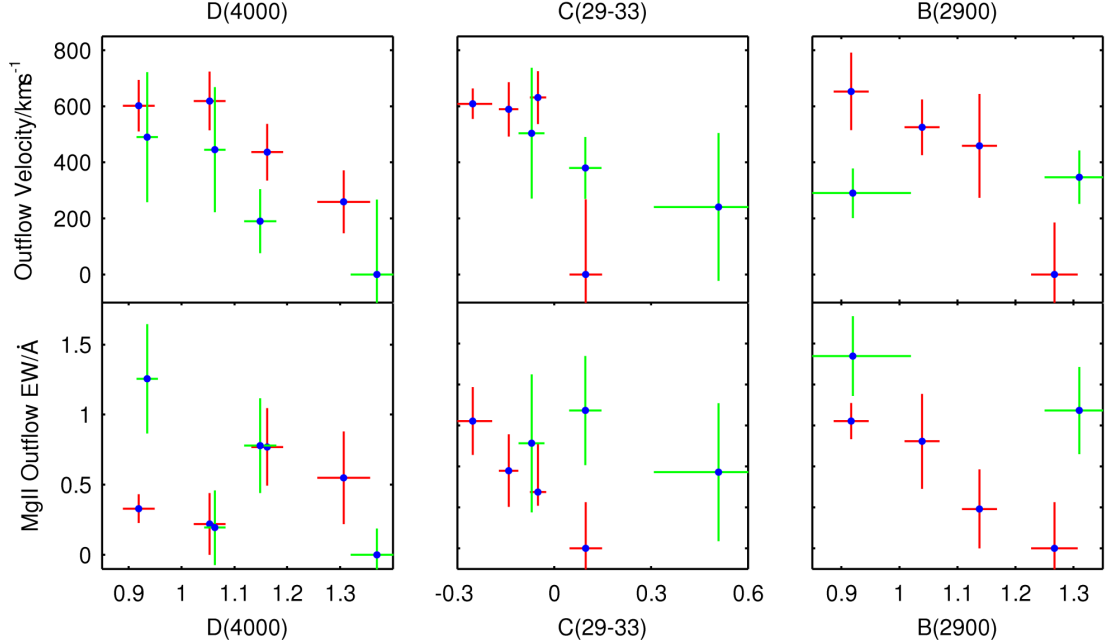} 
\end{center}
\caption{\small{This figure shows the velocity (top panels) and equivalent width (bottom panels) of the galactic outflows as a function of three spectral indices, as determined from two-component fits to the Mg\,\textsc{ii} absorption profile in composite spectra. The green points are based on the FORS2 sample, with the larger VIMOS sample shown in red. Data are plotted at the mean value of the spectral indices within each stack.}}
\label{fig:Fors2_spectral}
\end{figure*}

Figure~\ref{fig:Fors2_spectral} shows the velocities and equivalent widths for all three of the spectral indices defined, where the VIMOS data are plotted in red and the FORS2 data plotted in green. From this figure, we see evidence for higher velocity outflows among younger stellar populations in both the $D(4000)$ and $C(29-33)$ indices. The trend is also apparent in the $B(2900)$ index, but only for the VIMOS sample. Unfortunately the limited number of FORS2 spectra did not allow spectral stacks of sufficient S/N to probe the $B(2900)$ index in more than two bins.

None of the age-dependent indices show any trend with the equivalent width of the outflowing component in MgII. As mentioned previously, the equivalent width of MgII is difficult to interpret in stacked data (as the doublet is typically close to saturated) but this may indicate that the fraction of galaxies hosting an outflow is not strongly dependent on the age of the stellar populations.

\subsection{Rest-frame U-B Colour}\label{U-B}

We also use restframe $U-B$ colours to divide the galaxies into sub-samples. We use a colour-absolute magnitude diagram to make a visual distinction between the galaxies which lie on the `blue cloud' and those which lie on the `red sequence'. We define the blue cloud as galaxies with $U-B < 0.5$, which are generally found to be star-forming galaxies. Red sequence galaxies we define to have $U-B > 0.7$ and these are generally passive galaxies (although they can also be star-forming galaxies which are reddened by dust). The small population with a colour $0.7 > U-B > 0.5$ is classified as `green valley' galaxies.

\begin{figure*}
\begin{center}
\hspace*{-20pt}
\includegraphics[width=15.0cm]{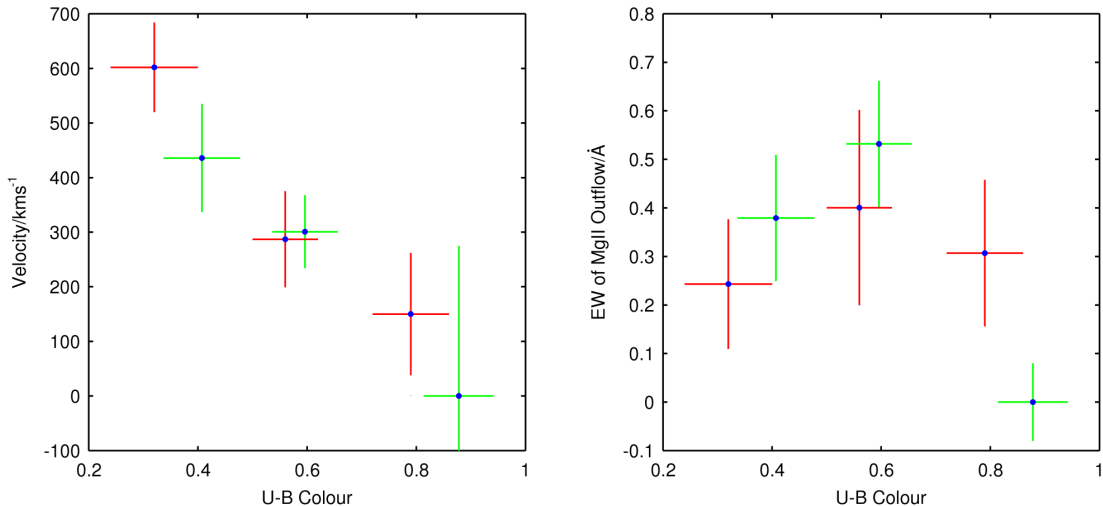} 
\end{center}
\caption{\small{This figure shows the outflow properties as a function of rest-frame $U-B$ colour. The left-hand panel shows the outflow velocities  determined from two-component fits to the composite spectra and the right-hand panel shows the equivalent widths associated with the outflows. Measurements from the VIMOS sample are shown in red, with the FORS2 measurements plotted in green. Data are plotted at the mean $U-B$ colour within each stack.}}
\label{fig:U-B}
\end{figure*}

The results of the two-component fitting analysis are summarised in Figure~\ref{fig:U-B}. It can be seen that the galaxy spectra show much the same correlation as has been seen in Section~\ref{definitions} and Section~\ref{OII_EW}. The galaxies in the blue cloud show the highest velocity outflows, whereas those on the red sequence show the lowest velocities.

Taken as a whole, the apparent trends of outflow velocity with [O\,\textsc{ii}] equivalent width, spectral indices and $U-B$ colour present very consistent trends; younger galaxies with enhanced star formation show the highest velocity outflows. Such trends strongly suggest that outflows are driven by stellar processes (e.g. stellar winds and supernovae), as discussed in \cite{2012ApJ...755L..26D}.

\section{Discussion}\label{termination}

From the outflow velocities we derive in this work, we can make a rough calculation of the amount of outflowing gas from the galaxies in our sample. Spectra of higher resolution reveal that the Mg\,\textsc{ii} absorption line is typically optically thick \citep{2009ApJ...692..187W}. To obtain an estimate of the column density of the outflowing gas we therefore use the method outlined in \cite{1968dms..book.....S}, using the ratio of the Mg\,\textsc{ii} doublet to estimate the optical depth. We cannot resolve the Mg\,\textsc{ii} doublet in our data and hence use the Mg\,\textsc{ii} column densities derived in this manner by \cite{2009ApJ...692..187W}, as the average properties of blue galaxies in both samples are very similar.

We transformed the estimates of the Mg\,\textsc{ii} column density into a hydrogen column density using a solar abundance ratio of log (Mg/H) = $-4.4$. The Mg depletion onto dust grains was then accounted for \citep[X = -1.2 dex,][]{{1996ARA&A..34..279S}} to obtain a hydrogen column density of $N(H) = 1.3 \times 10^{20}$ cm$^{-2}$.

Assuming the wind is a spherically symmetric thin shell of radius $R$, following \cite{2009ApJ...692..187W}, it can be shown that the mass outflow rate will be given by the following expression:
\begin{equation}
\dot{M} \simeq 44~M_{\rm \odot} \mathrm{yr}^{-1}~\frac{N_{H}}{10^{20}~\mathrm{cm}^{-2}} \frac{R}{5~\mathrm{kpc}} \frac{v}{600~\mathrm{kms}^{-1}}.
\end{equation}

The typical half-light radii of our galaxies are approximately 4~kpc. A diameter of 8~kpc also corresponds to the 1 arcsec slit width used for our spectroscopy. Assuming the outflows are of the order of the galaxy radius, for an outflow of velocity of 600~km\,s$^{-1}$ we obtain an outflow rate of approximately 45~$M_{\odot}$yr$^{-1}$. 

Our estimates for the outflow rate are subject to a number of uncertainties, particularly since we cannot resolve the Mg II doublet. It is notable, however, that assuming a high covering fraction, the implied mass outflow rates in the star-forming galaxies are similar to the star formation rates in these galaxies (10--50 $M_\odot$ yr$^{-1}$), suggesting that outflows may play a major role in processing the gas content of galaxies at these redshifts. This is consistent with the theory that the rate of gas inflowing into a galaxy is roughly equal to the gas processing rate, and of this gas, half is ejected in the form of outflows (as discussed in this work) and half is turned into stars. Such behaviour is consistent with models required to reproduce the high-redshift mass-metallicity relation, with star formation and outflows fed by the rapid accretion of external gas \citep[e.g.][]{{2008ApJ...674..151E}}.

We began this work with a discussion of AGN and starburst-driven winds and whether either mechanism could be responsible for terminating star formation at high redshift.  We established that high velocity outflows at $z>1$ are extremely common, and driven by starburst activity (due to the lack of AGN signatures in our composite spectra). The outflows in many of our stacks reach velocities approaching 1000\,km\,s$^{-1}$ (such as those from galaxies in the high SSFR and high-mass samples) are likely to exceed the escape velocities of their hosts. Our findings are consistent with the work of Diamond-Stanic et al. (2012), who suggest that galaxies with an extremely high star-formation rate surface density may have substantial momentum input from supernova and stellar winds to produce winds of even higher velocities ($>$1000\,km\,s$^{-1}$), without the need to invoke AGN-driven winds.

In this work, we see evidence that the age of the stellar population correlates strongly with outflow velocity. This is shown in Section~\ref{definitions}, where we investigate the wind properties of galaxies with varying spectral parameters. We find that the galaxies with younger stellar populations, as measured by the $D(4000)$ and $C(29-33)$ indices, as well as those galaxies with blue $U-B$ rest-frame colours and high SSFR, show the highest outflow velocities. Such trends may be explained by a time sequence, in which all galaxies go through a phase of hosting high-velocity outflows following a major burst of star formation, which subsequently slow down once the galaxies evolve to have older stellar populations.

In future work we will use morphological parameters to investigate the role of galaxy size and orientation on the observed outflows (e.g. by comparing edge-on and face-on systems). Orientation effects may easily dilute the true nature of the wind velocities in a stacking analysis. We can also use size information to investigate influence of the star-formation surface density (see Diamond-Stanic et al. 2012), to investigate if more compact star-formation at high redshift leads naturally to more intense outflows.

\section{Conclusions}\label{discussion}

In this work we use the spectra of 413 $K$-band selected galaxies to conduct a study of galactic-scale winds using a spectral stacking analysis at redshifts $0.71<z<1.63$. We use a combination of data from the VIMOS and FORS2 spectrographs, with the FORS2 spectra allowing us to probe older, redder galaxies at z $>$ 1, overcoming a limitation of previous work.  We stacked the spectra according to [O\,\textsc{ii}] equivalent width, star formation rate, galaxy mass, rest-frame colours and spectral indices.  We extract the velocities of the galactic winds by fitting a two-component model to the Mg\,\textsc{ii} absorption profiles, representing intrinsic and outflowing absorption.

Evidence for outflowing material is observed in virtually all stacks, suggesting that strong winds may be ubiquitous for star-forming galaxies at these redshifts, in agreement with previous findings \citep[e.g.][]{{2009ApJ...692..187W}}.  We find that the highest velocity outflows are hosted by blue galaxies with young stellar populations and strong [O\,\textsc{ii}] equivalent widths, with the highest velocity winds reaching velocities up to $\sim 1000$\,km\,s$^{-1}$.  This behaviour is consistent with outflows driven by stellar processes (e.g. stellar winds and supernovae).  The lack of AGN features in the composite spectra also suggest that AGN activity is not the primary energy source.

Our study was motivated by the findings of \cite{2007ApJ...663L..77T}, who observed very high velocity winds ( $>1000$\,km\,s$^{-1}$) from highly-luminous post-starburst galaxies at $z\sim 0.6$. Extrapolating from the trends observed in our study, we speculate that such extreme winds may be explained as the high-luminosity tail-end of the distribution for normal galaxies, without the need to invoke feedback from an AGN. 

Assuming a high covering fraction, the implied mass outflow rates in the star-forming galaxies are similar to the star formation rates in these galaxies ($10-50~M_{\odot}$yr$^{-1}$).  Such behaviour, if fed by the continuous accretion of cold gas, is consistent with models required to produce the mass-metallicity relation at high redshift \citep[e.g.][]{{2008ApJ...674..151E}}. 

\section*{Acknowledgments}

We are indebted to the staff at UKIRT for operating the telescope with such care and dedication. This work is partly based on observations made with the {\em Spitzer Space Telescope}, which is operated by the Jet Propulsion Laboratory, California Institute of Technology.  We also thank the teams at CASU and WFAU for processing and archiving the data. JSD acknowledges the support of the European Research Council via an Advanced Grant, and the support of the Royal Society through a Wolfson Research Merit Award.

\bibliographystyle{mn2e.bst}
\bibliography{Outflows}

\appendix

\section{Fors2}

\subsection{Methodology}\label{FORS2_method}

The FORS2 spectra have a resolution of $ \lambda / \Delta \lambda \sim$ 660 and therefore the Mg\,\textsc{ii} doublet at 2795.5, 2802.7 can be marginally separated, which is not the case in the VIMOS spectra. See Section 2 in the main text for further information on observational details of the FORS2 data.

We make a model Mg\,\textsc{ii} profile by convolving two lines with a ratio of 1.1:1 at 2795.5 \AA\, and 2802.7 \AA\, (to represent the absorption doublet) each with a Gaussian width of $\sigma= 60$\,km\,s$^{-1}$. The ratio represents the relative depths of the two lines in the Mg\,\textsc{ii} doublet if the absorption is optically thick, whilst the Gaussian represents an estimate of the average stellar velocity distribution, based on the work of \cite{2009ApJ...692..187W}. We then account for the resolution of the FORS2 instrument by further convolving the model Mg\,\textsc{ii} profile with another Gaussian with a FWHM of 5.0 \AA\, to represent the measured instrumental resolution, as determined from the strong [O\,\textsc{ii}] emission in the composite spectra.

The methodology for fitting the model Mg\,\textsc{ii} profile is then similar to to the procedures detailed in Section 3 of the main text with one major difference.  During the fitting procedure, approximately 15 pixels from the centre of the Mg\,\textsc{ii} profile in the composite spectrum are excluded from the fit. This is a consequence of being able to resolve the doublet due to the resolution of FORS2. The fitting algorithm produces a more stable fit if the pixels between the two minima in the Mg\,\textsc{ii} profile are removed. As before, however, the non-outflowing component is fixed at zero offset velocity, with the depth and velocity of the outflowing component allowed to vary. Errors are calculated using the same methodology as the errors used for the VIMOS spectra; they are the result of repeating each fit on 100 bootstrapped spectra.

\begin{figure*}
\begin{center}
\hspace*{-20pt}
\includegraphics[width=15.0cm]{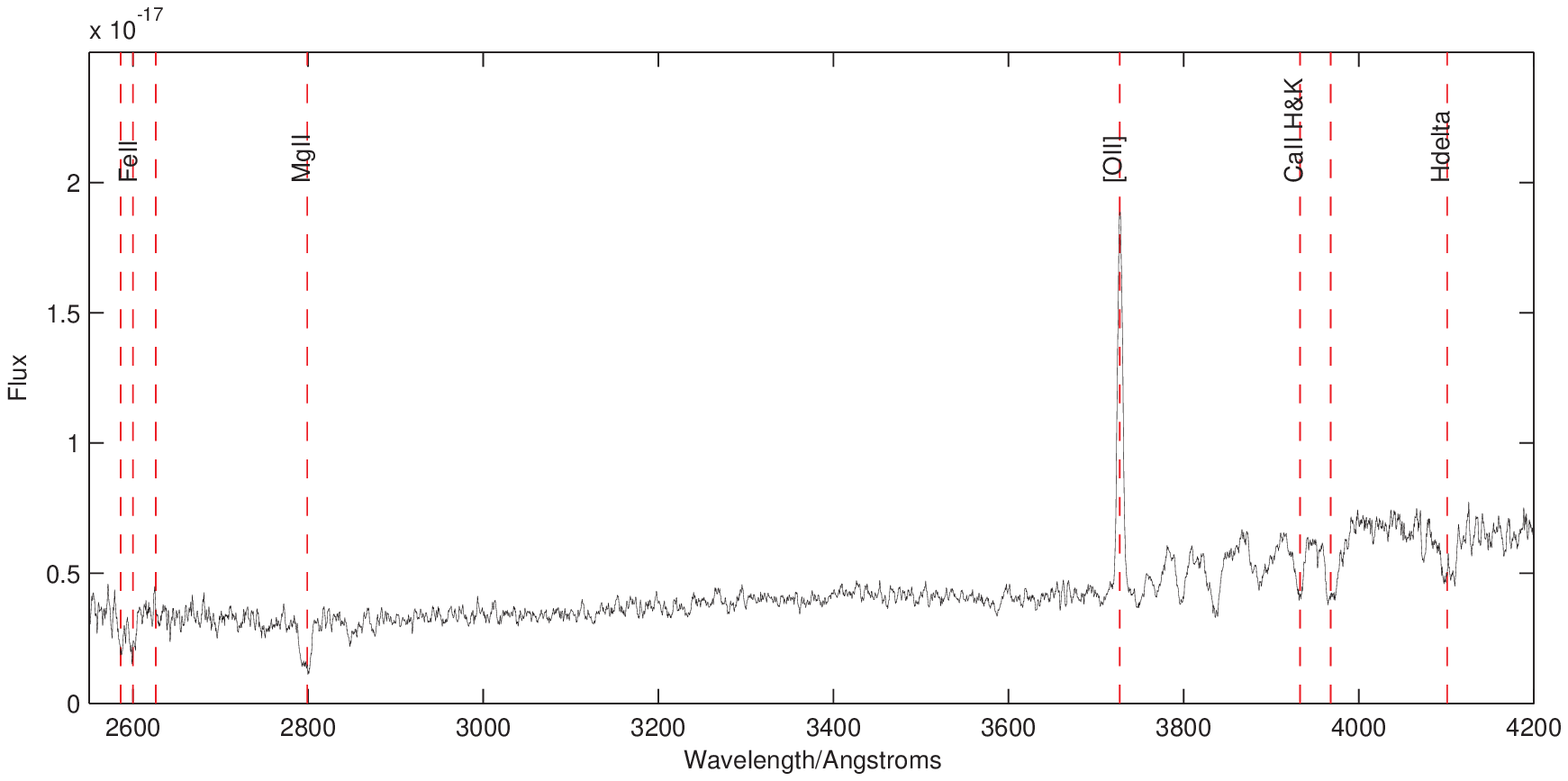} 
\end{center}
\caption{\small{The co-added stack of the 72 FORS2 spectra used in this work, all selected to ensure that  [O\,\textsc{ii}] and Mg\,\textsc{ii} were within the observed spectral range. Other emission and absorption features are also shown. Flux is shown in ergs~s$^{-1}$~cm$^{-2}$.}}
\label{fig:Fors2_full_spectra}
\end{figure*}

\begin{figure}
\begin{center}
\hspace*{-20pt}
\includegraphics[width=8.0cm]{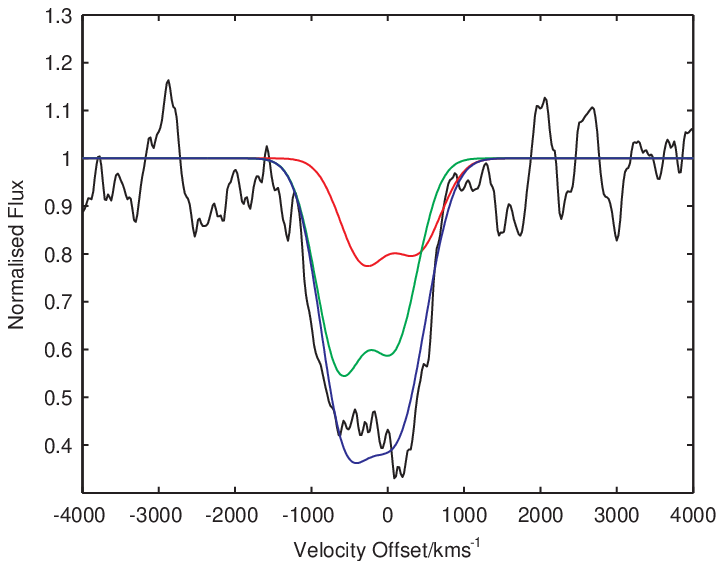} 
\end{center}
\caption{\small{This shows the Mg\,\textsc{ii} absorption profile from a co-added spectrum (black) containing all 72 FORS2 galaxies. We use a two-component fit to derive the outflowing component (green) and an intrinsic component in red (centred at 0 \,km\,s$^{-1}$). The blue line is the overall fit, which is the sum of the two components. The outflowing component is centred at 320\,km\,s$^{-1}$.}}
\label{fig:Fors2_zerovelocity}
\end{figure}

We create a composite of all FORS2 spectra (of which there are 72) in order to measure the overall velocity shift of the Mg\,\textsc{ii} profile in all galaxies and to see whether they agree with the results from the VIMOS data. The full wavelength range of the spectrum can be seen in Figure~\ref{fig:Fors2_full_spectra}. The Mg\,\textsc{ii} absorption from the composite spectrum is shown in Figure~\ref{fig:Fors2_zerovelocity}, and overplotted are the best fitting Mg\,\textsc{ii} profiles fitted using a two-component fit. The profile in green is the outflowing Mg\,\textsc{ii}, the red profile is the intrinsic Mg\,\textsc{ii} with its centre fixed at 0\,km\,s$^{-1}$. The blue profile is the sum of the two components and represents the overall fit to the absorption feature. Performing the two-component fit we measure an outflowing component with an offset by $\sim$320\,km\,s$^{-1}$, which is a similar to the velocity obtained from the VIMOS data ($\sim$340\,km\,s$^{-1}$).

\subsection{Outflows in Galaxies Stacked by Property}

We divide the FORS2 data into sub-samples and stack the data following the methodology in Section~\ref{stacking_method} in order to investigate and verify the trends found using the VIMOS data. The advantage of using the FORS2 data is that we can extend our analysis to those galaxies that are extremely red which are also at relatively high redshifts. In order to measure the velocity of the outflow in each composite, we use a method similar to that in Section~\ref{excess} in which two Mg\,\textsc{ii} profiles are fitted to the Mg\,\textsc{ii} absorption line.

\begin{figure*}
\begin{center}
\hspace*{-20pt}
\includegraphics[width=15.0cm]{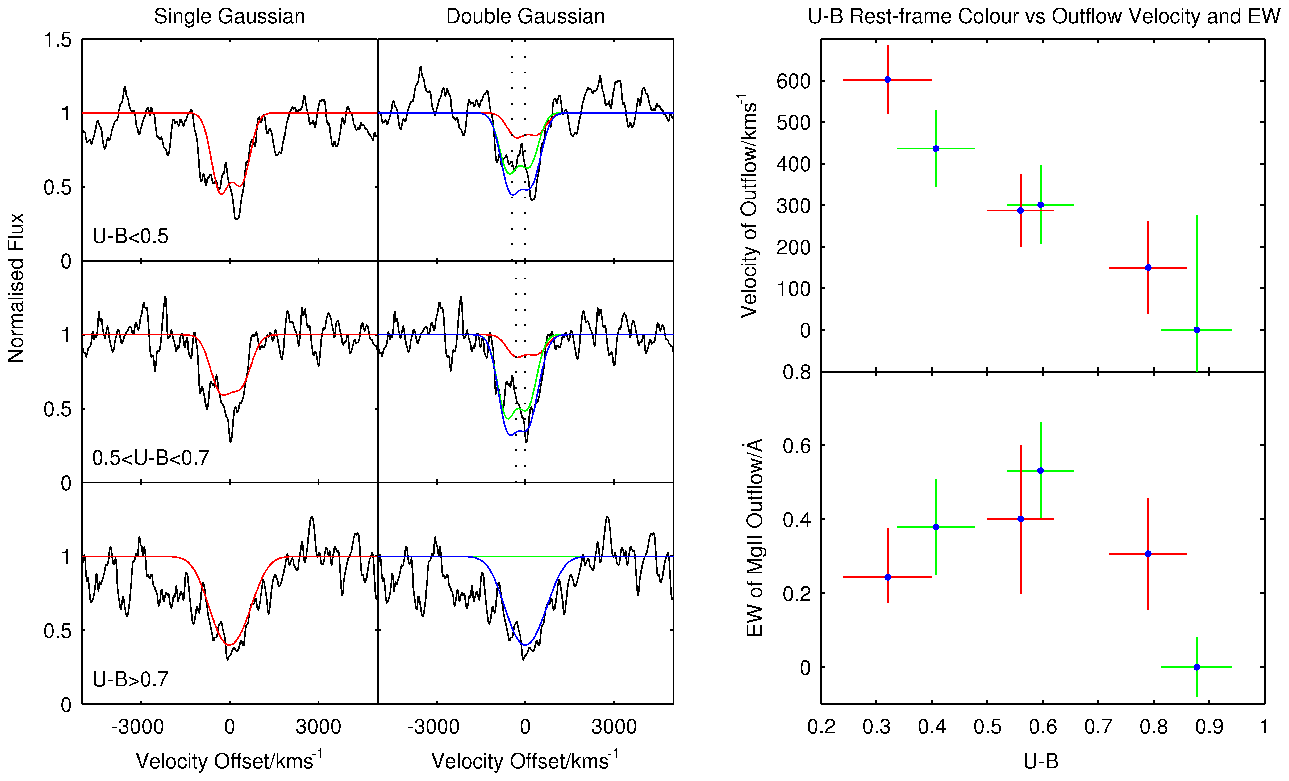} 
\end{center}
\caption{\small{This figure shows the three stacks of FORS2 spectra, categorised by their $U-B$ colour. The left-hand panels show the best fitting single Gaussian, fit to the red side of the Mg\,\textsc{ii} feature. The central column shows the breakdown of the absorption line into outflowing and non-outflowing components (green and red respectively), in addition to the overall best fit to the Mg\,\textsc{ii} line (in blue; the sum of the stationary and outflowing components). The upper-right panel shows the velocity of the outflowing component for each stack, and the lower right shows the equivalent width of this component (both in green), as compared to the equivalent VIMOS data, which is shown in red. From top to bottom, the stacks represent the `blue cloud', the `green valley', and the `red sequence' galaxies.}}
\label{fig:Fors2_UB}
\end{figure*}

We create the FORS2 composite spectra binned by the same parameters used for the VIMOS spectra; [O\,\textsc{ii}] equivalent width, $U-B$ colour, and spectral indices. The composites are not divided into the same sample bins as the VIMOS data, due to the differences in galaxy properties between the VIMOS and FORS2 samples and limitations in the number of galaxies per bin. For $U-B$ colour, we can create a red composite at $U-B >$ 0.7 with a significant number of spectra. We can create additional sub-samples for the C(29-33) parameter which extend to those galaxies which are extremely passive. The extracted velocities and equivalent widths from careful modelling of the Mg\,\textsc{ii} profiles can be seen in the main body of the text in Figures 4, 7 and 8 and these have been plotted against the results from the VIMOS data. In Figure~\ref{fig:Fors2_UB} we show an example of the Mg\,\textsc{ii} profile fitting with FORS2 data, using $U-B$ colour.

\bsp

\label{lastpage}

\end{document}